\documentclass[aps,prd,twocolumn,superscriptaddress]{revtex4-1}
\usepackage[utf8]{inputenc}
\usepackage[english]{babel}
\usepackage{bm}
\usepackage{graphicx}
\usepackage{epstopdf}
\usepackage{amsmath}
\usepackage{amssymb}
\usepackage{booktabs}
\usepackage{enumitem}
\usepackage{color}
\usepackage{hyperref}
\usepackage [autostyle, english = american]{csquotes}
\MakeOuterQuote{"}

\usepackage{mathtools} 	
\usepackage{comment} 	
\usepackage{multirow}	
\usepackage{gensymb}
\usepackage{epstopdf}
\usepackage[]{siunitx}  
\DeclareSIUnit\sample{S}
\DeclareSIUnit\bits{bits}

\usepackage{lineno} 	
\newcommand*\patchAmsMathEnvironmentForLineno[1]{%
  \expandafter\let\csname old#1\expandafter\endcsname\csname #1\endcsname
  \expandafter\let\csname oldend#1\expandafter\endcsname\csname end#1\endcsname
  \renewenvironment{#1}%
     {\linenomath\csname old#1\endcsname}%
     {\csname oldend#1\endcsname\endlinenomath}}%
\newcommand*\patchBothAmsMathEnvironmentsForLineno[1]{%
  \patchAmsMathEnvironmentForLineno{#1}%
  \patchAmsMathEnvironmentForLineno{#1*}}%
\AtBeginDocument{%
\patchBothAmsMathEnvironmentsForLineno{equation}%
\patchBothAmsMathEnvironmentsForLineno{align}%
\patchBothAmsMathEnvironmentsForLineno{flalign}%
\patchBothAmsMathEnvironmentsForLineno{alignat}%
\patchBothAmsMathEnvironmentsForLineno{gather}%
\patchBothAmsMathEnvironmentsForLineno{multline}%
}
\usepackage{xspace} 	

\makeatletter
\renewcommand\tableofcontents{%
    \@starttoc{toc}%
}
\makeatother

\newif\ifshowchanges
\showchangesfalse

\newcommand{\vo}[1]{}

\ifshowchanges
\usepackage{soul}
\soulregister\cite7
\soulregister\ref7
	\renewcommand{\vo}[1]{{\textcolor{red}{\st{#1}}}}
    
\fi

\usepackage{babel}

\begin{document}

\title{Searches for Double Beta Decay of $^{134}\text{Xe}$ with EXO-200}
\collaboration{EXO-200 Collaboration}

\author{J.B.~Albert}\affiliation{Physics Department and CEEM, Indiana University, Bloomington, Indiana 47405, USA}
\author{G.~Anton}\affiliation{Erlangen Centre for Astroparticle Physics (ECAP), Friedrich-Alexander-University Erlangen-N\"urnberg, Erlangen 91058, Germany}
\author{I.~Badhrees}\altaffiliation{Permanent with King Abdulaziz City for Science and Technology, Riyadh, Saudi Arabia}\affiliation{Physics Department, Carleton University, Ottawa, Ontario K1S 5B6, Canada}
\author{P.S.~Barbeau}\affiliation{Department of Physics, Duke University, and Triangle Universities Nuclear Laboratory (TUNL), Durham, North Carolina 27708, USA}
\author{R.~Bayerlein}\affiliation{Erlangen Centre for Astroparticle Physics (ECAP), Friedrich-Alexander-University Erlangen-N\"urnberg, Erlangen 91058, Germany}
\author{D.~Beck}\affiliation{Physics Department, University of Illinois, Urbana-Champaign, Illinois 61801, USA}
\author{V.~Belov}\affiliation{Institute for Theoretical and Experimental Physics, Moscow, Russia}
\author{M.~Breidenbach}\affiliation{SLAC National Accelerator Laboratory, Menlo Park, California 94025, USA}
\author{T.~Brunner}\affiliation{Physics Department, McGill University, Montr\'eal, Qu\'ebec H3A 2T8, Canada}\affiliation{TRIUMF, Vancouver, British Columbia V6T 2A3, Canada}
\author{G.F.~Cao}\affiliation{Institute of High Energy Physics, Beijing, China}
\author{W.R.~Cen}\affiliation{Institute of High Energy Physics, Beijing, China}
\author{C.~Chambers}\affiliation{Physics Department, Colorado State University, Fort Collins, Colorado 80523, USA}
\author{B.~Cleveland}\affiliation{Department of Physics, Laurentian University, Sudbury, Ontario P3E 2C6, Canada}\affiliation{SNOLAB, Sudbury, Ontario P3Y 1N2, Canada}
\author{M.~Coon}\affiliation{Physics Department, University of Illinois, Urbana-Champaign, Illinois 61801, USA}
\author{A.~Craycraft}\affiliation{Physics Department, Colorado State University, Fort Collins, Colorado 80523, USA}
\author{W.~Cree}\affiliation{Physics Department, Carleton University, Ottawa, Ontario K1S 5B6, Canada}
\author{T.~Daniels}\affiliation{SLAC National Accelerator Laboratory, Menlo Park, California 94025, USA}
\author{M.~Danilov}\altaffiliation{Now at P.N.Lebedev Physical Institute of the Russian Academy of Sciences, Moscow, Russia}\affiliation{Institute for Theoretical and Experimental Physics, Moscow, Russia}
\author{S.J.~Daugherty}\affiliation{Physics Department and CEEM, Indiana University, Bloomington, Indiana 47405, USA}
\author{J.~Daughhetee}\affiliation{Physics Department, University of South Dakota, Vermillion, South Dakota 57069, USA}
\author{J.~Davis}\affiliation{SLAC National Accelerator Laboratory, Menlo Park, California 94025, USA}
\author{S.~Delaquis}\affiliation{SLAC National Accelerator Laboratory, Menlo Park, California 94025, USA}
\author{A.~Der~Mesrobian-Kabakian}\affiliation{Department of Physics, Laurentian University, Sudbury, Ontario P3E 2C6, Canada}
\author{R.~DeVoe}\affiliation{Physics Department, Stanford University, Stanford, California 94305, USA}
\author{T.~Didberidze}\affiliation{Department of Physics and Astronomy, University of Alabama, Tuscaloosa, Alabama 35487, USA}
\author{J.~Dilling}\affiliation{TRIUMF, Vancouver, British Columbia V6T 2A3, Canada}
\author{A.~Dolgolenko}\affiliation{Institute for Theoretical and Experimental Physics, Moscow, Russia}
\author{M.J.~Dolinski}\affiliation{Department of Physics, Drexel University, Philadelphia, Pennsylvania 19104, USA}
\author{W.~Fairbank Jr.}\affiliation{Physics Department, Colorado State University, Fort Collins, Colorado 80523, USA}
\author{J.~Farine}\affiliation{Department of Physics, Laurentian University, Sudbury, Ontario P3E 2C6, Canada}
\author{S.~Feyzbakhsh}\affiliation{Amherst Center for Fundamental Interactions and Physics Department, University of Massachusetts, Amherst, MA 01003, USA}
\author{P.~Fierlinger}\affiliation{Technische Universit\"at M\"unchen, Physikdepartment and Excellence Cluster Universe, Garching 80805, Germany}
\author{D.~Fudenberg}\affiliation{Physics Department, Stanford University, Stanford, California 94305, USA}
\author{R.~Gornea}\affiliation{Physics Department, Carleton University, Ottawa, Ontario K1S 5B6, Canada}\affiliation{TRIUMF, Vancouver, British Columbia V6T 2A3, Canada}
\author{K.~Graham}\affiliation{Physics Department, Carleton University, Ottawa, Ontario K1S 5B6, Canada}
\author{G.~Gratta}\affiliation{Physics Department, Stanford University, Stanford, California 94305, USA}
\author{C.~Hall}\affiliation{Physics Department, University of Maryland, College Park, Maryland 20742, USA}
\author{J.~Hoessl}\affiliation{Erlangen Centre for Astroparticle Physics (ECAP), Friedrich-Alexander-University Erlangen-N\"urnberg, Erlangen 91058, Germany}
\author{P.~Hufschmidt}\affiliation{Erlangen Centre for Astroparticle Physics (ECAP), Friedrich-Alexander-University Erlangen-N\"urnberg, Erlangen 91058, Germany}
\author{M.~Hughes}\affiliation{Department of Physics and Astronomy, University of Alabama, Tuscaloosa, Alabama 35487, USA}
\author{A.~Jamil}\affiliation{Erlangen Centre for Astroparticle Physics (ECAP), Friedrich-Alexander-University Erlangen-N\"urnberg, Erlangen 91058, Germany}\affiliation{Physics Department, Stanford University, Stanford, California 94305, USA}
\author{M.J.~Jewell}\affiliation{Physics Department, Stanford University, Stanford, California 94305, USA}
\author{A.~Johnson}\affiliation{SLAC National Accelerator Laboratory, Menlo Park, California 94025, USA}
\author{S.~Johnston}\altaffiliation{Now at Argonne National Laboratory, Argonne, Illinois USA}\affiliation{Amherst Center for Fundamental Interactions and Physics Department, University of Massachusetts, Amherst, MA 01003, USA}
\author{A.~Karelin}\affiliation{Institute for Theoretical and Experimental Physics, Moscow, Russia}
\author{L.J.~Kaufman}\affiliation{Physics Department and CEEM, Indiana University, Bloomington, Indiana 47405, USA}
\author{T.~Koffas}\affiliation{Physics Department, Carleton University, Ottawa, Ontario K1S 5B6, Canada}
\author{S.~Kravitz}\affiliation{Physics Department, Stanford University, Stanford, California 94305, USA}
\author{R.~Kr\"{u}cken}\affiliation{TRIUMF, Vancouver, British Columbia V6T 2A3, Canada}
\author{A.~Kuchenkov}\affiliation{Institute for Theoretical and Experimental Physics, Moscow, Russia}
\author{K.S.~Kumar}\affiliation{Department of Physics and Astronomy, Stony Brook University, SUNY, Stony Brook, New York 11794, USA}
\author{Y.~Lan}\affiliation{TRIUMF, Vancouver, British Columbia V6T 2A3, Canada}
\author{D.S.~Leonard}\affiliation{IBS Center for Underground Physics, Daejeon, Korea}
\author{S.~Li}\affiliation{Physics Department, University of Illinois, Urbana-Champaign, Illinois 61801, USA}
\author{C.~Licciardi}\affiliation{Physics Department, Carleton University, Ottawa, Ontario K1S 5B6, Canada}
\author{Y.H.~Lin}\affiliation{Department of Physics, Drexel University, Philadelphia, Pennsylvania 19104, USA}
\author{R.~MacLellan}\affiliation{Physics Department, University of South Dakota, Vermillion, South Dakota 57069, USA}
\author{M.G.~Marino}\affiliation{Technische Universit\"at M\"unchen, Physikdepartment and Excellence Cluster Universe, Garching 80805, Germany}
\author{T.~Michel}\affiliation{Erlangen Centre for Astroparticle Physics (ECAP), Friedrich-Alexander-University Erlangen-N\"urnberg, Erlangen 91058, Germany}
\author{B.~Mong}\affiliation{SLAC National Accelerator Laboratory, Menlo Park, California 94025, USA}
\author{D.~Moore}\affiliation{Department of Physics, Yale University, New Haven, Connecticut 06511, USA}
\author{K.~Murray}\affiliation{Physics Department, McGill University, Montr\'eal, Qu\'ebec H3A 2T8, Canada}
\author{R.~Nelson}\affiliation{Waste Isolation Pilot Plant, Carlsbad, New Mexico 88220, USA}
\author{O.~Njoya}\affiliation{Department of Physics and Astronomy, Stony Brook University, SUNY, Stony Brook, New York 11794, USA}
\author{A.~Odian}\affiliation{SLAC National Accelerator Laboratory, Menlo Park, California 94025, USA}
\author{I.~Ostrovskiy}\affiliation{Department of Physics and Astronomy, University of Alabama, Tuscaloosa, Alabama 35487, USA}
\author{A.~Piepke}\affiliation{Department of Physics and Astronomy, University of Alabama, Tuscaloosa, Alabama 35487, USA}
\author{A.~Pocar}\affiliation{Amherst Center for Fundamental Interactions and Physics Department, University of Massachusetts, Amherst, MA 01003, USA}
\author{F.~Reti\`{e}re}\affiliation{TRIUMF, Vancouver, British Columbia V6T 2A3, Canada}
\author{P.C.~Rowson}\affiliation{SLAC National Accelerator Laboratory, Menlo Park, California 94025, USA}
\author{J.J.~Russell}\affiliation{SLAC National Accelerator Laboratory, Menlo Park, California 94025, USA}
\author{A.~Schubert}\affiliation{Physics Department, Stanford University, Stanford, California 94305, USA}
\author{D.~Sinclair}\affiliation{Physics Department, Carleton University, Ottawa, Ontario K1S 5B6, Canada}\affiliation{TRIUMF, Vancouver, British Columbia V6T 2A3, Canada}
\author{E.~Smith}\affiliation{Department of Physics, Drexel University, Philadelphia, Pennsylvania 19104, USA}
\author{V.~Stekhanov}\affiliation{Institute for Theoretical and Experimental Physics, Moscow, Russia}
\author{M.~Tarka}\affiliation{Department of Physics and Astronomy, Stony Brook University, SUNY, Stony Brook, New York 11794, USA}
\author{T.~Tolba}\affiliation{Institute of High Energy Physics, Beijing, China}
\author{R.~Tsang}\altaffiliation{Now at Pacific Northwest National Laboratory, Richland, Washington, USA}\affiliation{Department of Physics and Astronomy, University of Alabama, Tuscaloosa, Alabama 35487, USA}
\author{P.~Vogel}\affiliation{Kellogg Lab, Caltech, Pasadena, California 91125, USA}
\author{J.-L.~Vuilleumier}\affiliation{LHEP, Albert Einstein Center, University of Bern, Bern, Switzerland}
\author{M.~Wagenpfeil}\affiliation{Erlangen Centre for Astroparticle Physics (ECAP), Friedrich-Alexander-University Erlangen-N\"urnberg, Erlangen 91058, Germany}
\author{A.~Waite}\affiliation{SLAC National Accelerator Laboratory, Menlo Park, California 94025, USA}
\author{J.~Walton}\affiliation{Physics Department, University of Illinois, Urbana-Champaign, Illinois 61801, USA}
\author{T.~Walton}\affiliation{Physics Department, Colorado State University, Fort Collins, Colorado 80523, USA}
\author{M.~Weber}\affiliation{Physics Department, Stanford University, Stanford, California 94305, USA}
\author{L.J.~Wen}\affiliation{Institute of High Energy Physics, Beijing, China}
\author{U.~Wichoski}\affiliation{Department of Physics, Laurentian University, Sudbury, Ontario P3E 2C6, Canada}
\author{L.~Yang}\affiliation{Physics Department, University of Illinois, Urbana-Champaign, Illinois 61801, USA}
\author{Y.-R.~Yen}\affiliation{Department of Physics, Drexel University, Philadelphia, Pennsylvania 19104, USA}
\author{O.Ya.~Zeldovich}\affiliation{Institute for Theoretical and Experimental Physics, Moscow, Russia}
\author{J.~Zettlemoyer}\affiliation{Physics Department and CEEM, Indiana University, Bloomington, Indiana 47405, USA}
\author{T.~Ziegler}\affiliation{Erlangen Centre for Astroparticle Physics (ECAP), Friedrich-Alexander-University Erlangen-N\"urnberg, Erlangen 91058, Germany}

\date{\today}
\begin{abstract}
Searches for double beta decay of $^{134}\text{Xe}$ were performed
with EXO-200, a single-phase liquid xenon detector designed to search
for neutrinoless double beta decay of $^{136}\text{Xe}$. Using an
exposure of $29.6\text{ kg}\!\cdot\!\text{yr}$, the lower limits
of $\text{T}_{1/2}^{2\nu\beta\!\beta}>8.7\cdot10^{20}\text{ yr}$
and $\text{T}_{1/2}^{0\nu\beta\!\beta}>1.1\cdot10^{23}\text{ yr}$
at 90\% confidence level were derived, with corresponding half-life
sensitivities of $1.2\cdot10^{21}\text{ yr}$ and $1.9\cdot10^{23}\text{ yr}$.
These limits exceed those in the literature for $^{134}\text{Xe}$, improving
by factors of nearly $10^{5}$ and 2 for the two antineutrino and neutrinoless modes,
respectively.
\end{abstract}
\maketitle

\section{\label{sec:introduction}Introduction}

This paper presents the search for two modes of double beta ($\beta\!\beta$)
decay of $^{134}\text{Xe}$. $\beta\!\beta$ decay is a second-order
weak transition between two nuclei with the same mass number and nuclear
charges that differ by two units. This process can only be observed
if the single beta ($\beta$) decay is strongly suppressed or forbidden
by energy conservation. The mode with emission of two antineutrinos
and two electrons ($2\nu\beta\!\beta$) is an allowed decay by the
Standard Model (SM) and has been directly observed in nine nuclei~\cite{pdg}.
Among them, $^{136}\text{Xe}$ presents the longest half-life 
of $2.165\pm0.016\,\text{(stat.)}\pm0.059\,\text{(syst.)}\cdot10^{21}\text{ yr}$~\cite{bb2n-precise}. The hypothetical
neutrinoless mode with emission of two electrons and nothing else
($0\nu\beta\!\beta$) does not conserve lepton number and, if observed,
would imply that neutrinos are massive Majorana particles \cite{theorem}.
The most stringent lower limits derived for the half-life of $0\nu\beta\!\beta$
in $^{136}\text{Xe }$ 
are $1.1\cdot10^{26}\text{ yr}$~\cite{klz} and 
$1.1\cdot10^{25}\text{ yr}$~\cite{bb0n-nature} 
at 90\% confidence level
(CL).

The $\beta\!\beta$ decay of $^{134}\text{Xe}$ into $^{134}\text{Ba}$:
\begin{equation*}
^{134}\text{Xe}\rightarrow^{134}\text{Ba}^{++}+2\,e^{-}\left(+2\,\bar{\nu}_{e}\right)~,\label{eq:xe134-bb2n}
\end{equation*}
has a Q-value of $825.8\pm0.9$ keV \cite{xe134-qval} and neither
of the two $\beta\!\beta$ modes have been observed to date. 
Because $\beta\!\beta$ decay rates scale strongly with the Q-value: $\sim Q^{11}$ in $2\nu\beta\!\beta$
and $\sim Q^{5}$ in $0\nu\beta\!\beta$~\cite{q-scale2,q-scale}, 
experimental searches have favored $^{136}\text{Xe}$ (Q-value of $2457.83\pm0.37\text{ keV}$~\cite{xe136-qval}).
Moreover, in xenon detectors
containing both isotopes, $^{136}\text{Xe}$ $2\nu\beta\!\beta$ produces a background
that makes the $\beta\!\beta$ searches in $^{134}\text{Xe}$ even
more challenging. The current experimental limit for the half-life of $2\nu\beta\!\beta$
in $^{134}\text{Xe}$ is $\text{T}_{1/2}^{2\nu\beta\!\beta}>1.1\cdot10^{16}\text{ yr}$
at 68\% CL~\cite{xe134-bb2n}, while theoretical predictions put it in the
range of $\sim10^{24}-10^{25}\text{ yr}$ \cite{xe134-overview}.
On the other hand, 
more recent searches set the most stringent limit for the $0\nu\beta\!\beta$
half-life at $\text{T}_{1/2}^{0\nu\beta\beta}>5.8\cdot10^{22}\text{ yr}$
at 90\% CL \cite{xe134-bb0n}.

The searches presented in this paper are rooted in the success of
the EXO-200 analyses of $\beta\!\beta$ decays in $^{136}\text{Xe}$~\cite{bb2n-precise,bb0n-nature,majoron,lorentz}.
Unique to this work, the energy threshold was extended to lower energies as required
by the $\beta\!\beta$ searches in $^{134}\text{Xe}$. 
As will be discussed in Sec. \ref{sec:analysis}, each decay mode was analyzed independently, 
using a different energy threshold. 
The Monte Carlo (MC) simulation 
and reconstruction processes were improved, as detailed in Sec. \ref{sec:detector},
to further improve the agreement between data and MC. 
Another change
with respect to previous EXO-200 analyses is the use of the full set
of data between June 2011, and February 2014, 
corresponding to a 25\% increase in livetime (EXO-200 Phase-I).

\section{\label{sec:detector}The EXO-200 Detector, Data and MC Simulation}

The EXO-200 detector consists of two back-to-back cylindrical single-phase
time projection chambers (TPCs), sharing a central cathode, filled
with liquid xenon (LXe). The isotopic composition of the LXe is 
$80.672\pm0.014\%$ $^{136}\text{Xe}$ and $19.098\pm0.014\%$ $^{134}\text{Xe}$.
The ratio between these two isotopes was measured using dynamic dual-inlet mass spectrometry~\cite{isotopic}.
 In addition, the contamination from other Xe isotopes was measured to be $<0.25\%$, 
dominated by a $0.2\%$ contamination of $^{132}\text{Xe}$.
The significant concentration of $^{134}\text{Xe}$, almost twice its
natural abundance of 10.4\%, presents a unique opportunity 
and motivates this work. 

The detector is located at the Waste Isolation Pilot Plant (WIPP) in
Carlsbad, NM, USA, in a clean room under an overburden of 1624 meters
water equivalent. An active muon veto system surrounding the clean
room on four sides identifies 96\% of the cosmic ray muons passing
through the TPCs, and allows rejection of prompt cosmogenic backgrounds
\cite{cosmogenics}.

A radiopure copper vessel, nearly $44\text{ cm}$ in length and $40\text{ cm}$
in diameter, contains the EXO-200 TPCs. Each TPC is instrumented near
the ends of the vessel with a pair of wire planes, crossed at $60^{\circ}$,
in front of an array of silicon large-area avalanche photodiodes (APDs). Ionizing 
particles passing through the LXe deposit energy that produces both scintillation
light ($\sim178\text{ nm}$ wavelength), detected by the APDs almost instantenously,
and electron-ion pairs. The electrons are drifted towards the wire
grids, inducing signals in the front-most wire plane (V-wires), and
then are collected by the second wire plane (U-wires). 
Copper "field-shaping" rings ensure a sufficient uniformity of the electric field over
the bulk of the LXe, and inside them 
a cylindrical PTFE reflector improves collection efficiency of the
scintillation light. A more detailed description of the detector can
be found in~\cite{jinst}.

All three spatial coordinates (3D) of the energy depositions are reconstructed
in EXO-200. Information from the U- and V-wires results in 2D clusters ($X$
and $Y$ coordinates) formed by the charge detection (charge clusters).
The time difference between the light signal (scintillation cluster)
and associated charge clusters provides their third coordinate ($Z$).
The subcentimeter position resolution~\cite{bb2n-precise} provides 
strong separation between single-site (SS) 
events, primarily $\beta$ or $\beta\!\beta$ decays with characteristic dimension of $\sim2-3\text{ mm}$, 
and multi-site (MS) 
events, arising mostly from multiple interactions 
of MeV-energy $\gamma$-rays.
Furthermore, internally generated $\beta$-like events in the fiducial volume (FV) are uniformly distributed in the
LXe, in contrast to the spatial distribution of background events arising from
$\gamma$-rays entering the TPC. This difference is captured in the analysis by
the standoff-distance variable, defined as the shortest distance between
any event cluster and the closest material that is not LXe, other
than the cathode. The event energy is calculated using a linear combination of
the measured ionization and scintillation signals that optimizes the energy resolution~\cite{prb}, 
determined using the $2615\text{ keV}$ $\gamma$-line of $^{208}\text{Tl}$.

Both the spectral fitting analysis, presented in Sec. \ref{sec:analysis}, and detector calibration
rely on detailed modeling of the detector response. For these purposes, a GEANT4-based application~\cite{geant4} 
is part of the EXO-200 MC
simulation software, as described in~\cite{bb2n-precise}. The collaboration has been
gradually implementing changes into this package to better describe
the measurements with the detector \cite{majoron}. For
this analysis, the simulation was updated to incorporate three important effects,
in order to improve the spectral agreement with data at low energies. 
First, since electronegative impurities can capture electrons drifting in the LXe,
the charge collection is exponentially attenuated with drifting distance
before the simulation of the electronics pulse shapes.
The average electron
lifetime included in the simulation is based on 
calibration measurements ($\bar{\tau}_{e}=4.5\text{ ms}$). 
In addition, a more realistic light response
of the APDs is included, which is now based on EXO-200 data to account for the
complexity of optical propagation in the detector, such as internal reflections. 
Finally, the diffusion
of the drifting electrons has been incorporated following the EXO-200
measurement of the transverse coefficient in LXe at the nominal drift
field of $380\text{ V/cm}$ ($D=55\text{ cm}^{2}\text{/s}$)~\cite{diffusion}.

The energy calibration relies on data acquired with radioactive $\gamma$
sources deployed near the detector \cite{bb2n-precise}. 
The energy scale and resolution are simultaneously determined by fitting
the expected energy spectra, as generated by MC, to the corresponding
calibration data~\cite{majoron}. 
These fits were performed with a reduced
energy threshold suited for both $\beta\!\beta$ decay searches of
$^{134}\text{Xe}$. The effective livetime-weighted average of the resolution in this analysis
is $\sigma/E=$ 1.60\% and 3.56\% for SS events at the
Q-value of $^{136}\text{Xe}$ and $^{134}\text{Xe}$, respectively. 
To reach this result, a sophisticated de-noising algorithm was developed optimizing
the energy resolution in the presence of correlated noise from the APD electronics~\cite{denoise}.

The total livetime of the EXO-200 data considered here is 596.7 days. The
fiducial volume (FV) is defined by events within $10\text{ mm}<|Z|<182\text{ mm}$,
where $Z=0$ is the cathode plane, and constrained in a hexagon with
$162\text{ mm}$ apothem, centered at $(X,Y)=(0,0)$. This corresponds
to $18.1\text{ kg}$ of $^{134}\text{Xe}$, \emph{i.e.} $8.14\cdot10^{25}$
atoms, which results in an exposure of $29.6$ kg$\cdot$yr ($221$ mol$\cdot$yr).

\section{\label{sec:analysis}Analysis Procedure}

\begin{figure}[h]
\noindent \begin{centering}
\includegraphics[scale=0.38]{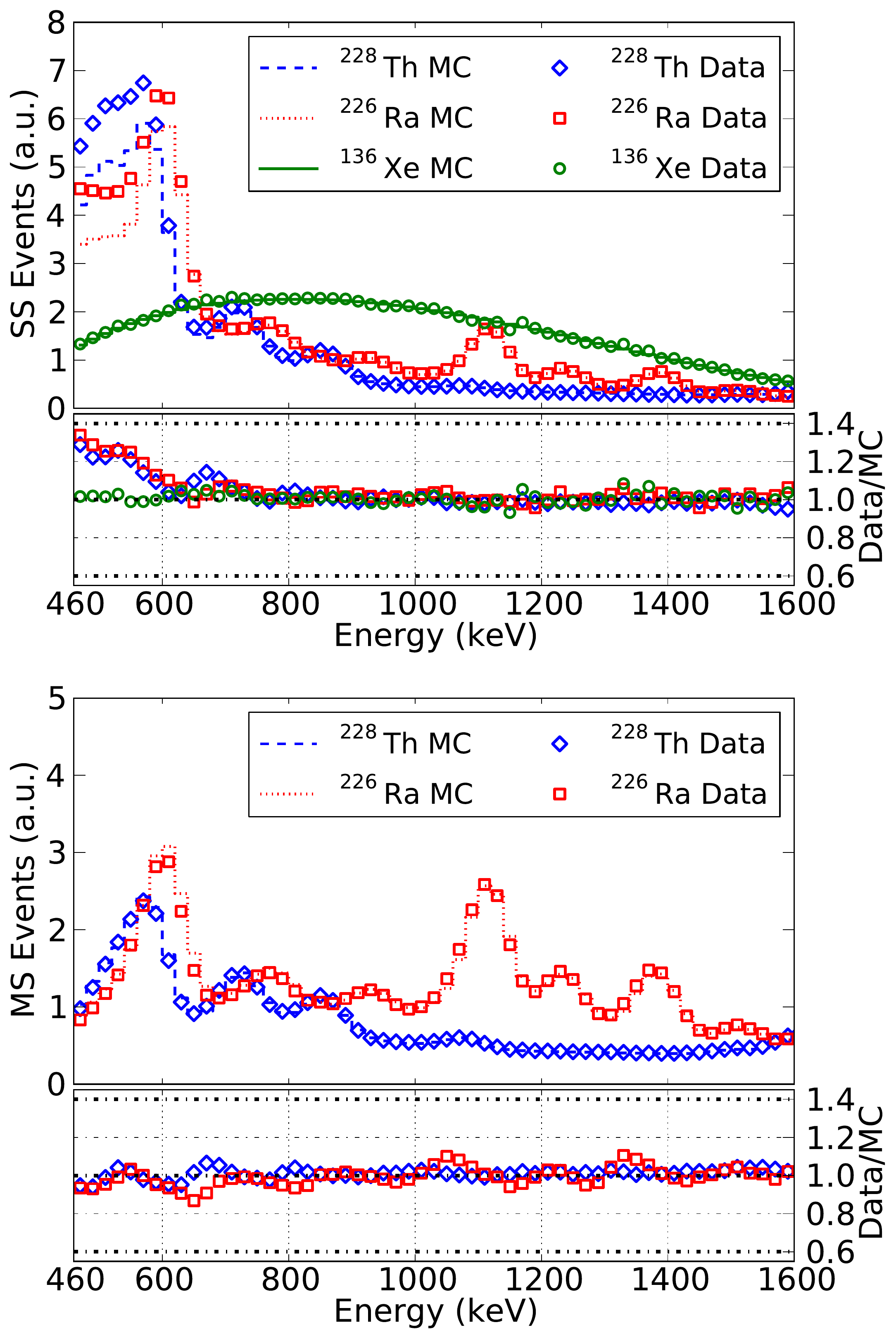}
\par\end{centering}
\caption{\label{fig:source-agree}Energy spectral agreement between data and
MC simulation in SS (top) and MS (bottom) events. The comparison is presented 
between 460 keV and 1600 keV
for two calibration sources, $^{228}\text{Th}$ and $^{226}\text{Ra}$, deployed near the cathode of the detector
for calibration runs. The background-subtracted 
$^{136}\text{Xe-}2\nu\beta\!\beta$ spectrum is compared to MC only in SS events.
A worsening of the agreement in SS $\gamma$-like spectra is observed for energies $\lesssim600\text{ keV}$. 
This effect was considered in the searches as discussed in Sec. \ref{sec:errors}.}
\end{figure}

The low Q-value of the $\beta\!\beta$ decay of $^{134}\text{Xe}$
requires an energy threshold that is substantially lower than the 980~keV
used in other EXO-200 publications. The improvements described in
Sec. \ref{sec:detector} 
produce an agreement between data and MC better than $10\%$
for energies above 600~keV,
as shown in Fig. \ref{fig:source-agree}.
The agreement worsens below this energy and reaches 30\% near $460\text{ keV}$
in SS events induced by $\gamma$-rays 
from calibration sources. 
The effects of these discrepancies are 
discussed in Sec. \ref{sec:errors}.
The standoff-distance
agreement was not observed to degrade at low energies when compared to previous EXO-200 analyses. 

Following a similar procedure from previous analyses~\cite{bb0n-nature,bb2n-precise,lorentz,majoron},
the search for each $\beta\!\beta$ decay mode of $^{134}\text{Xe}$ was performed independently using
a binned negative log-likelihood (NLL) function to fit simultaneously both SS and MS events with
their corresponding probability density functions (PDFs), as generated by MC, in energy and standoff-distance. 
Five Gaussian constraints, presented in Sec. \ref{sec:errors}, were included in the NLL function
to incorporate the systematic uncertainties independently evaluated for each search, in a similar approach as~\cite{bb2n-precise}.
The SS fraction of each component, defined
by the ratio of the number of SS events to the total number of events
(SS/(SS+MS)), parametrizes the proportion of counts assigned
to SS and MS PDFs. Unlike previous analyses, these searches used non-uniform
bin widths, which optimize the calculation
speed without decreasing the experimental sensitivity. In particular,
the standoff-distance binning partitions the LXe in equal volumes.
A profile-likelihood scan was then performed to derive the limits at 90\% CL 
using a profile-likelihood ratio ($\Delta\text{NLL}$) of 1.35, under the assumption of Wilks's theorem~\cite{wilks,analysis},
which applies, given the large statistics of the data set in the region of interest.

A fit model comprising the significant components that contribute 
to events with energies above $700\text{ keV}$
was developed in~\cite{bb2n-precise}.
At lower energies, two additional components are expected to contribute to backgrounds:
\begin{itemize}
\item $^{85}\text{Kr}$ dissolved in the LXe, producing $\beta$ decays with end point
at $687.4\text{ keV}$; 
\item $^{137}\text{Cs}$ in the materials near the LXe, with $\gamma$-rays
of $661.7\text{ keV}$.
\end{itemize}

The shape of the simulated $\beta\!\beta$ decay spectrum of $^{134}\text{Xe}$ is
the same as that of $^{136}\text{Xe}$, with the appropriate
Q-value. The simulated energy spectrum of $^{85}\text{Kr}$ 
includes the two $\beta$-decay modes with branching ratios of 99.56\%
and 0.44\% to the ground and excited states of $^{85}\text{Rb},$
respectively. The latter is followed by the release of a 
$514\text{ keV}$ $\gamma$-ray. 
A shape correction accounting for the forbidden nature of the 
first unique $\beta$ decay was calculated 
using the method described in~\cite{kr85},
and found between $-15\%$ and $80\%$ depending on its energy.
This correction was applied as an event weighting in the MC simulation.

The possible difference between the energy scale of $\beta$- and $\gamma$-like events is
modeled by a scaling factor, the $\beta$-scale. This is a free parameter
applied on the $\beta$-like PDFs that 
allows for a possible shift in energy scale between $\beta$-like PDFs and the $\gamma$ calibration sources.

Different energy thresholds are used to optimize 
the sensitivities for $2\nu\beta\!\beta$ and $0\nu\beta\!\beta$ decays.
The $2\nu\beta\!\beta$ decay requires the lowest possible energy threshold,
in order to maximize the signal detection efficiency and 
discrimination power between low-energy backgrounds, while keeping the
systematic errors arising from the spectral agreement at reasonable
levels. Because all these effects are propagated into the profile-likelihood
ratio, the sensitivity (obtained through fits of toy data sets 
generated by the background model) was evaluated with energy thresholds varying
between $400\text{ keV}$ and $500\text{ keV}$, in steps of $20\text{ keV}$,
and found to be optimal in the region between $460\text{ keV}$ and $480\text{ keV}$ 
(with negligible differences within this range).
The choice of 460 keV 
can then be motivated by its signal detection efficiency, 5.6\% as opposed to 4.5\%. 
On the other hand, the $0\nu\beta\!\beta$ detection efficiency 
is nearly maximal, 89\%, for all energies below $760\text{ keV}$. 
The energy threshold of this search is then selected at $740\text{ keV}$,
sufficiently away from the low-energy background components, even 
when accounting for the energy resolution.
For this reason, the $^{85}\text{Kr}$ and $^{137}\text{Cs}$
PDFs are only included in the fit model of the 
$^{134}\text{Xe}$ $2\nu\beta\!\beta$ search.

\section{\label{sec:errors}Systematic Errors}

The five Gaussian constraints added to the NLL, responsible for the propagation of the 
systematic errors into the searches, correspond to: 
\begin{itemize}
\item uncertainty in the activity of radon in the LXe as determined in stand-alone studies;
\item uncertainty in the relative fractions of neutron-capture-related PDF components generated
by dedicated simulations; 
\item uncertainty in SS fractions as obtained in MC;
\item uncertainty in the overall efficiency, also referred to as \emph{normalization}, caused by 
imperfections in the MC model;
\item uncertainty in the signal efficiency, also referred to as \emph{signal-specific normalization}, caused by  
spectral differences between data and MC simulations. 
\end{itemize}
The first two were evaluated in previous analyses~\cite{bb2n-precise}, 
and are presented in Table \ref{tab:systs} along with the other three, explained below.
Table \ref{tab:contribs} shows the contribution of each constraint to the 90\% CL
limits (derived in Sec. \ref{sec:results} and shown in Fig. \ref{fig:profiles}), evaluated by setting a negligible error
to the constraint in the fit.

The uncertainty in SS fractions was evaluated using calibration data 
and was defined as the weighted average of the
SS fractions residuals ((data-MC)/MC), with weights based on the signal
spectrum and detector livetime. 
Since the SS fraction is observed to depend on energy, 
being $\gtrsim90\%$ for energies below 700 keV
for all components, this error was considered as the largest 
between those evaluated in energy and standoff-distance projections.
The resulting SS-fraction
constraint for each search is shown in Table \ref{tab:systs}.

Imperfections in the MC model, common to all components, translate
into an overall difference in number of events between data and MC
prediction. This overall efficiency uncertainty is acccounted for by an additional degree
of freedom added to the fitting PDF through a normalization parameter
that scales all PDF coefficients equally. This normalization
factor is constrained to unity within the estimated systematic error,
whose largest contributions arise from the FV cut and the 3D clustering
step of the reconstruction \cite{bb0n-nature}. Using a similar approach as in previous
analyses, these errors were found to be 5.8\% (3.6\%) and 2.3\% (3.1\%), respectively,
for the $2\nu\beta\!\beta$ ($0\nu\beta\!\beta$) analysis. Other sources were found to
contribute negligibly ($\lesssim1\%$) to the total normalization
error, shown in Table \ref{tab:systs}.

Discrepancies in the shape distributions between data and MC are propagated
into the signal rate through a normalization parameter that only scales
the coefficient of the signal PDFs. This signal-specific normalization
parameter is constrained to unity within the errors arising from spectral
shape agreement and background model. 

To estimate the effect of shape errors,
the ratio between data and MC of the projections onto energy, shown
in Fig. \ref{fig:source-agree}, 
and standoff-distance were used to weight all PDF components
(also referred to as \emph{un-skewing}).
The standoff-distance ratios, as well as those for energies above 850~keV, 
were found to be negligible contributors.
$^{60}\text{Co}$ and $^{238}\text{U}$-related
PDFs were weighted by ratios using data from the calibration sources
$^{60}\text{Co}$ and $^{226}\text{Ra}$, respectively, while the
other $\gamma$-like PDFs were weighted by ratios obtained with a $^{228}\text{Th}$
source. $\beta$-like PDFs were weighted using ratios
from the background subtracted $^{136}\text{Xe}$-$2\nu\beta\beta$ spectrum (Fig.~\ref{fig:source-agree}),
since these are also uniformly distributed in LXe. Approximately 10,000 toy datasets 
were drawn from these un-skewed PDFs, which were 
scaled by values arising from a data fit by the background-only model (without a signal PDF),
while the number of signal counts included in the toy fits was manually set. 
These toy data sets were then fit using
the normal PDFs and the average difference (bias) between the manually set and best-fit
number of signal counts determined. 
This bias was found to be roughly constant
at $2250\text{ cts}$ ($240\text{ cts}$) for the $2\nu\beta\!\beta$
($0\nu\beta\!\beta$) analysis. 
The difference between these factors demonstrates the impact of the
spectral discrepancy at low energies. 

The dependence of the signal rate on the completeness of the fit model
was studied by individually including possible background contributors
in different locations and/or from other decays. The relative change
of the estimated rate was then determined. This change was found to be negligible
for $^{39}\text{Ar}$ and $^{42}\text{Ar}$
dissolved in the LXe, and for $^{60}\text{Co}$ and $^{238}\text{U}$ in components 
farther than the TPC vessel.
The dominant contribution to this term in the $2\nu\beta\!\beta$ search arose from $^{210}\text{Bi}$ (10\%),
while in the $0\nu\beta\!\beta$ search 
$^{85}\text{Kr}$ in the LXe (12\%) dominated. The
impact of the $^{85}\text{Kr}$ on the $2\nu\beta\!\beta$ search is discussed in Sec. \ref{sec:results}.

The total deviations arising from background model uncertainties are shown in
Table \ref{tab:systs} ($a$), along with the estimated errors from the spectral agreement ($b$).
The signal-specific normalization error is the largest systematic contribution in both searches,
as presented in Table \ref{tab:contribs}, contributing to 34.6\% (30.4\%)  increase of the
90\% CL limit derived for  $^{134}\text{Xe}$ $2\nu\beta\!\beta$ ($0\nu\beta\!\beta$).

\begin{table}[h]
\caption{\label{tab:systs}Summary of the constraints added to
the searches of $\beta\!\beta$ decays in $^{134}\text{Xe}$.
The signal-specific normalization error is calculated
by $\sigma=\sqrt{(a\cdot n)^2+b^2}$, where $n$ is the number of signal counts.
}
\noindent \begin{centering}
\begin{tabular}{c|c|c}
\hline 
Constraint & $2\nu\beta\!\beta$ & $0\nu\beta\!\beta$\tabularnewline
\hline 
Radon in the LXe & 10\% & 10\%\tabularnewline
Neutron-capture PDF fractions & 20\% & 20\%\tabularnewline
SS fractions & 5.7\% & 2.3\%\tabularnewline
Normalization & 6.2\% & 4.9\%\tabularnewline
Signal-specific normalization & $\begin{array}{c}
a=11.8\%\;\\
b=2250\text{ cts}
\end{array}$ & $\begin{array}{c}
a=12.7\%\;\\
b=240\text{ cts}
\end{array}$\tabularnewline
\hline 
\end{tabular}
\par\end{centering}
\end{table}

\begin{table}[h]
\caption{\label{tab:contribs}Contribution of each systematic error to the 90\% CL limits derived in
the searches of $^{134}\text{Xe}$ $\beta\!\beta$-decays, presented in Sec. \ref{sec:results}.}
\noindent \begin{centering}
\begin{tabular}{c|c|c}
\hline 
Error Contribution & $2\nu\beta\!\beta$ & $0\nu\beta\!\beta$\tabularnewline
\hline 
Radon in the LXe & $<0.1\%$ & $<0.1\%$ \tabularnewline
Neutron-capture PDF fractions & $<0.1\%$ & $<0.1\%$ \tabularnewline
SS fractions & 16.6\% & 10.2\%\tabularnewline
Normalization & 1.0\% & 0.2\%\tabularnewline
Signal-specific normalization & 34.6\% & 30.4\% \tabularnewline
\hline 
\end{tabular}
\par\end{centering}
\end{table}

\section{\label{sec:results}Results and Discussion}

\begin{figure}[h]
\noindent \begin{centering}
\includegraphics[scale=0.37]{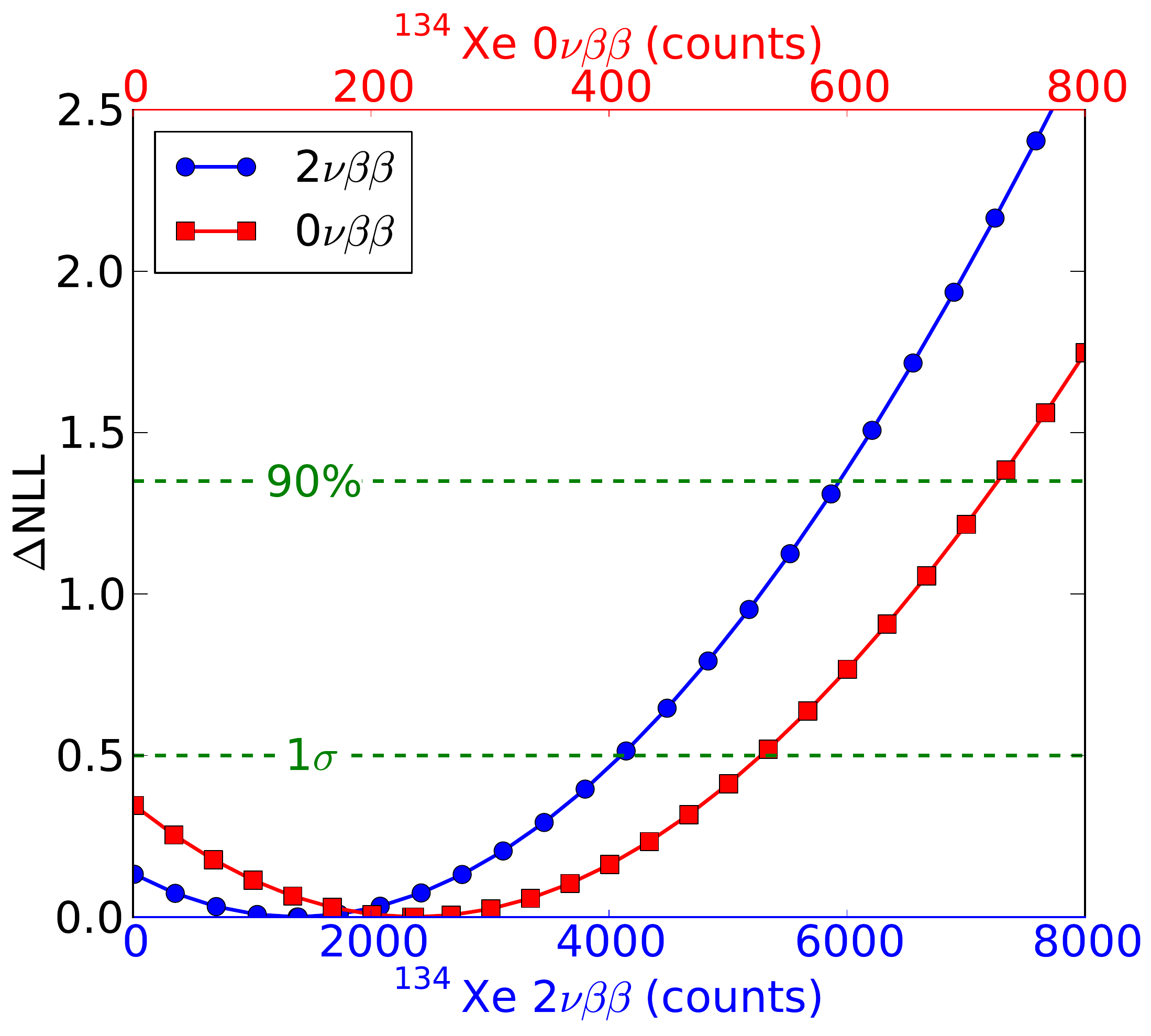}
\par\end{centering}
\caption{\label{fig:profiles}Profile-likelihood ratio, $\Delta\text{NLL}$,
for $^{134}\text{Xe}$ $2\nu\beta\!\beta$ and $0\nu\beta\!\beta$.
The dashed lines represent the $1\sigma$ and 90\% CLs, assuming the
validity of Wilks's theorem \cite{wilks,analysis}. The latter intersects
the profile curves at 5900 and 730 counts for $2\nu\beta\!\beta$ and
$0\nu\beta\!\beta$, respectively.}
\end{figure}

\begin{figure*}[!]
\noindent \begin{centering}
\includegraphics[scale=0.55]{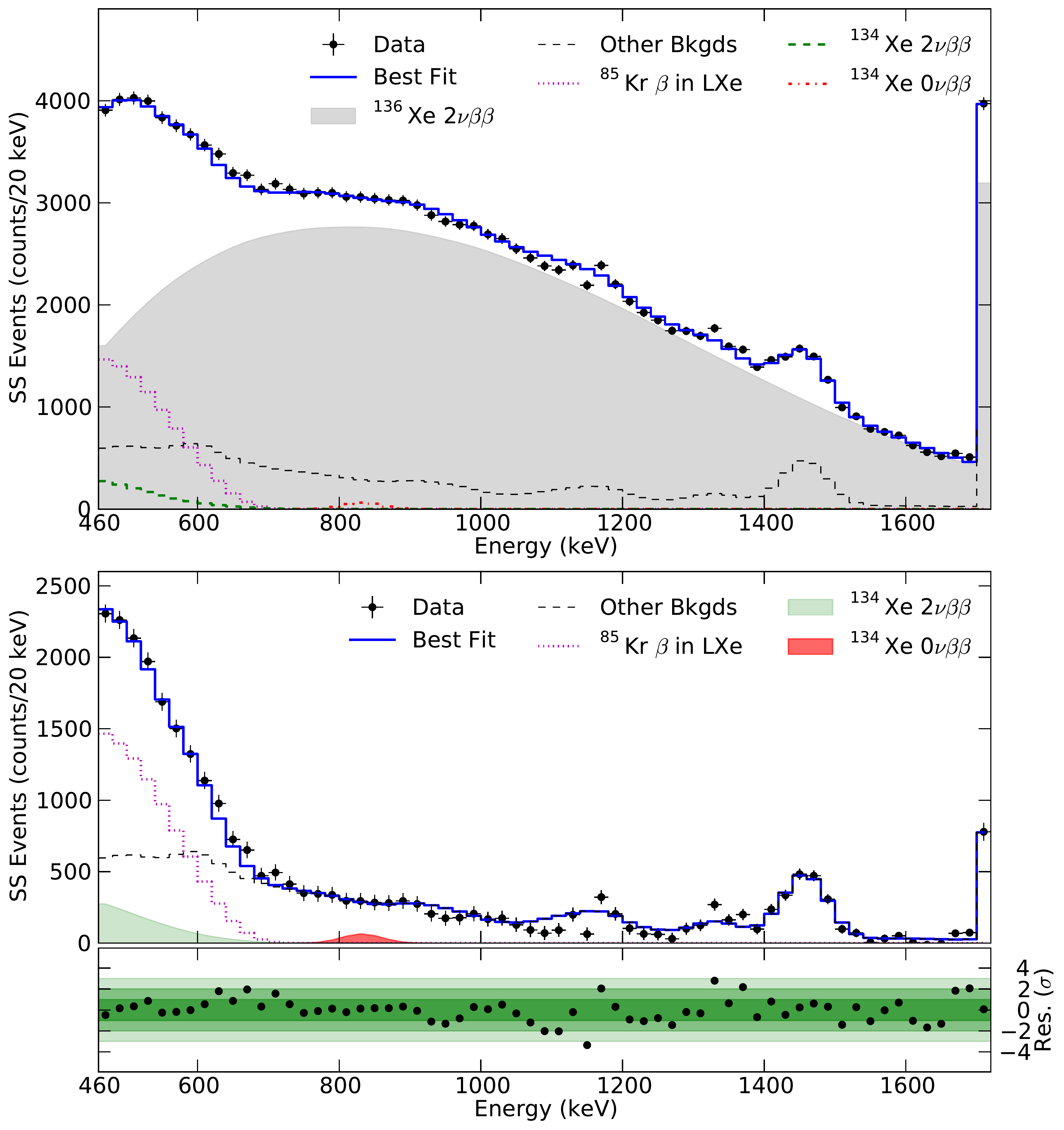}
\par\end{centering}
\caption{\label{fig:fits} The top plot shows the SS energy projection for the
results of the $^{134}\text{Xe}$-$2\nu\beta\!\beta$
analysis, using the fitted central values. 
While the $0\nu\beta\!\beta$ search
is performed in a separate fit with higher energy threshold,
it is overlaid here for comparison since the backgrounds
at these energies are the same. 
The last bin is the summed contents of all bins between 1700 keV and 9800 keV.
The middle plot shows the same results where the $^{136}\text{Xe}$-$2\nu\beta\!\beta$
component is subtracted from data and best fit values. The bottom plot presents
the residuals between data and best fit normalized to the Poisson error.}
\end{figure*}

Figure \ref{fig:profiles} shows the profile-likelihood scan performed
for the $^{134}\text{Xe}$ $2\nu\beta\!\beta$ and $0\nu\beta\!\beta$ decays, where the lower limits of 
$\text{T}_{1/2}^{2\nu\beta\!\beta}>8.7\cdot10^{20}\text{ yr}$
and $\text{T}_{1/2}^{0\nu\beta\!\beta}>1.1\cdot10^{23}\text{ yr}$
at 90\% CL are derived for their half-lives, respectively.
The corresponding experimental sensitivities were evaluated at $1.2\cdot10^{21}\text{ yr}$
and $1.9\cdot10^{23}\text{ yr}$, respectively.
The results of the NLL fit from the $^{134}\text{Xe}$ $2\nu\beta\!\beta$ 
search are presented in Fig. \ref{fig:fits}, along with the fitted $^{134}\text{Xe}$ $0\nu\beta\!\beta$ 
from the other search. The limits presented in this paper
increase the sensitivity relative to those available in the current literature
by 5 orders of magnitude for the $2\nu\beta\!\beta$ search~\cite{xe134-bb2n},
while the limit set for $0\nu\beta\!\beta$ is
nearly twice as stringent than the one in~\cite{xe134-bb0n}.

The significance of the presence of a signal relative to the null hypothesis
is calculated using fits of toy data sets and comparing the NLL between hypotheses. 
The \emph{p}-values were found to be 0.24 and 0.19 for the 
$2\nu\beta\!\beta$ and $0\nu\beta\!\beta$ searches, respectively, 
showing that there is no statistically significant evidence for a non-zero signal.

Both fitted $\beta$-scales
are consistent with unity to the subpercent level. The fitted half-life
of $2\nu\beta\!\beta$ of $^{136}\text{Xe}$ agrees to better than $1\%$
with its precise measurement in~\cite{bb2n-precise}, which was obtained
with a different analysis on a subset of the present data. The observed
residuals from both analyses are comparable to those in \cite{bb0n-nature},
performed with a reduced data set. 

\begin{figure}[tb]
\noindent \begin{centering}
\includegraphics[scale=0.37]{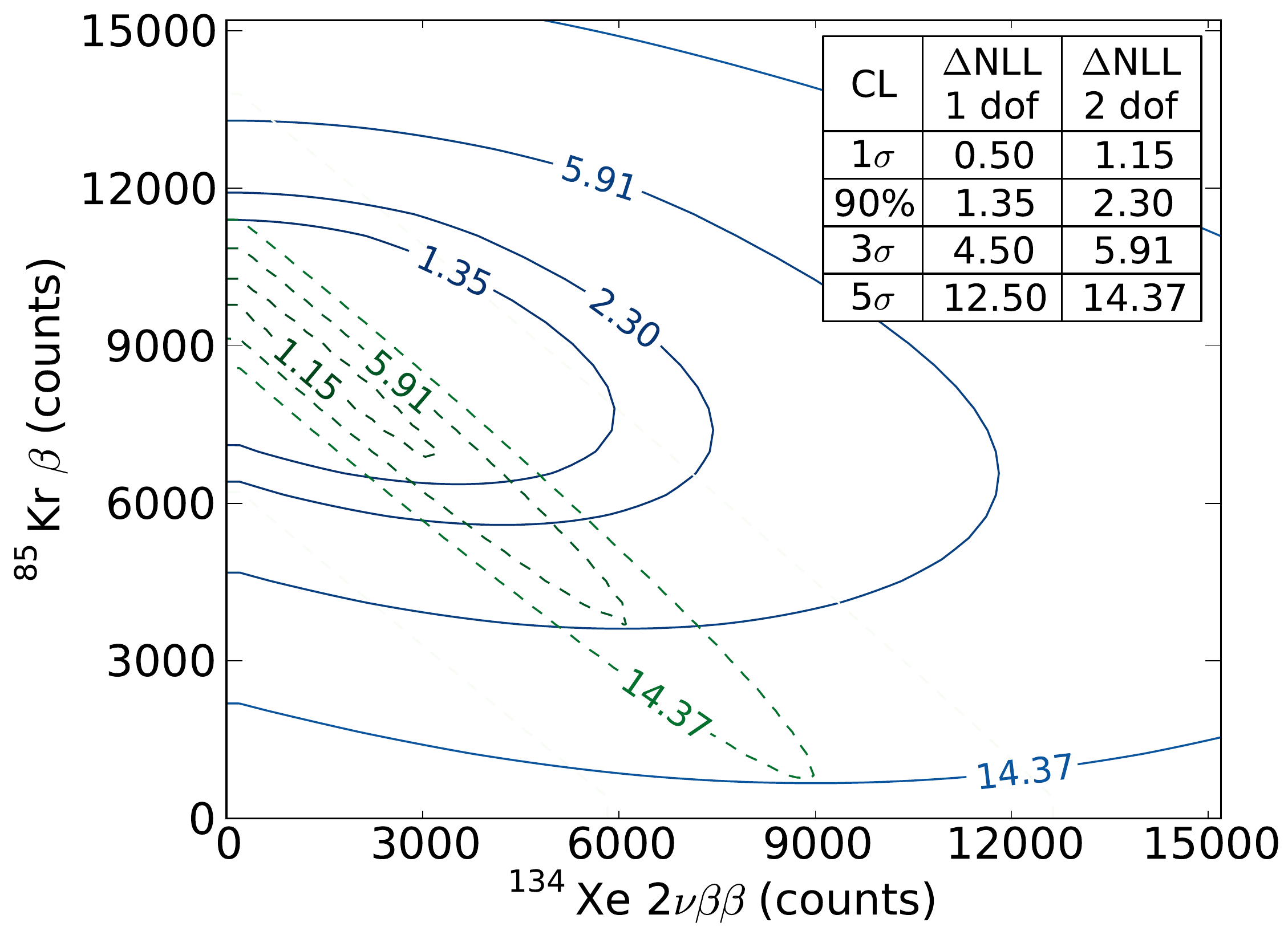}
\par\end{centering}
\caption{\label{fig:correlation}Contour lines of the profile-likelihood ratio, $\Delta \text{NLL}$,
scanned for the $^{85}\text{Kr}$ $\beta$ and $^{134}\text{Xe}$ $2\nu\beta\!\beta$ components
using two NLL functions. The blue solid lines were obtained
using the same function as the $^{134}\text{Xe}$ $2\nu\beta\!\beta$ search,
which accounts for all the systematic errors. The green dashed lines were evaluated without consideration
of the normalization errors. The table shows $\Delta$NLL values of various CL for 1 and 2 dof.} 
\end{figure}

Figure \ref{fig:correlation} shows the 
contour lines of the profile-likelihood ratio scanned
for $^{85}\text{Kr}$ $\beta$ and $^{134}\text{Xe}$ $2\nu\beta\!\beta$.
The solid lines were evaluated incorporating 
all the systematic errors in the NLL function, exactly as in the $^{134}\text{Xe}$ 
$2\nu\beta\!\beta$ search, whereas the dashed lines were obtained
without consideration of either the normalization error nor the signal-specific one. 
The latter better represents the large correlation between the two PDFs, $\rho=-0.97$, 
as expected from their shapes presented in Fig. \ref{fig:fits},
and shows the extent to which they are not fully degenerate. Contour lines for 2 degrees of freedom (dof)
were evaluated, showing the regions where both variables
are contained with corresponding CL. 
 The data prefer non-zero $^{85}\text{Kr}$ $\beta$ counts,
which was checked to be driven by SS shape differences. As a result,
the upper limit set on $^{134}\text{Xe}$ $2\nu\beta\!\beta$, 5900 cts, is improved
by a factor of nearly two, if compared to the limit that would be set by the sum of components, 
estimated by the intersection of the solid $\Delta\text{NLL}=2.30$ line
and the y-axis, 12000 cts.

A contamination of
$25.5\pm3\text{ ppt}$ natural krypton in the enriched LXe 
was measured before filling the TPC~\cite{exo-nim}. Combined
with the current work, which does not perform a complete measurement of the $^{85}\text{Kr}$ contamination,
these results suggest an isotopic abundance in the enriched LXe
consistent with those at atmospheric levels, ${\sim10^{-11}\text{ g }^{85}\text{Kr/g }^{\text{nat.}}\text{Kr}}$~\cite{kr85-conc}.
Since the right edge of the solid lines can be identified with the profile depicted in Fig. \ref{fig:profiles},
contour lines for 1 dof were also evaluated in this case. Thus, the impact of $^{85}\text{Kr}$ $\beta$ in the $^{134}\text{Xe}$-$2\nu\beta\!\beta$ limit
can be estimated by its difference to the limit that would be set for 
a fixed amount of $^{85}\text{Kr}$ $\beta$. Considering this value to be that near the NLL minimum, $9000$ cts,
the solid $\Delta\text{NLL}=1.35$ line in Fig. \ref{fig:correlation} indicates a contribution of about $15\%$
increase in the $^{134}\text{Xe}$-$2\nu\beta\!\beta$ 90\% CL upper limit. 
Further, as might be expected given the very different energy response,
the impact of this uncertainty on the $0\nu\beta\!\beta$ search limit is significantly smaller.

EXO-200 has begun Phase-II data taking, after a two-year hiatus,
with upgraded electronics that may result
in better detection efficiency at low energies as well as improved
spectral agreement between data and MC simulation. These improvements can
positively impact future EXO-200 searches for $\beta\!\beta$-decay of
$^{134}\text{Xe}$. In the long term, the proposed nEXO detector is
projected to increase the EXO-200 sensitivity to $0\nu\beta\!\beta$
in $^{136}\text{Xe}$ by nearly 3 orders of magnitude~\cite{nexo}.
While the sensitivity for $\beta\!\beta$-decay in $^{134}\text{Xe}$ has
not been directly studied yet,
a similar increase in performance for $^{134}\text{Xe}$ would allow
this next generation experiment to probe the $2\nu\beta\!\beta$
decay of this isotope to half-lives within the theoretical expectations.

\begin{acknowledgments}
The EXO-200 collaboration acknowledges the KARMEN collaboration for
supplying the cosmic-ray veto detectors, and the WIPP for their hospitality.
EXO-200 is supported by DOE and NSF in the United States, NSERC in
Canada, SNF in Switzerland, IBS in Korea, RFBR in Russia,
DFG Cluster of Excellence ``Universe'' in Germany, and CAS and ISTCP 
in China. EXO-200 data analysis and
simulation uses resources of the National Energy Research Scientific
Computing Center (NERSC). 
\end{acknowledgments}

\end{document}